\documentclass[showpacs,aps,graphicx]{revtex4}
\usepackage{graphicx}

\begin{document}

\title{Cascaded  noiseless linear  amplification for single-photon state }

\author{Lan Zhou$^{1,2}$, Yu-Bo Sheng$^{2,3,\ast}$ }
\address{$^1$ College of Mathematics \& Physics, Nanjing University of Posts and Telecommunications, Nanjing,
210003, China\\
$^2$ Key Lab of Broadband Wireless Communication and Sensor Network
 Technology, Nanjing University of Posts and Telecommunications, Ministry of
 Education, Nanjing, 210003, China\\
 $^3$Institute of Signal Processing  Transmission, Nanjing
University of Posts and Telecommunications, Nanjing, 210003,  China\\}

\date{\today}

\begin{abstract}
Photon loss is one of the main obstacles in current long-distance quantum communications. The approach of  noiseless linear amplification (NLA) is one of the powerful way to distill the single-photon state (SPS) from a mixed state, which comprises both the SPS and vacuum state. However, existing NLA protocol
can only perform the amplification for one time. That is the fidelity of the SPS cannot be increased anymore. In this paper,
We put forward an efficient cascaded NLA protocol for both the SPS and single-photon entanglement (SPE), respectively, with the help of some auxiliary single photons. By repeating this protocol for sever times, the fidelity of the SPS and SPE can reach near 100\%, which may make this protocol is extremely useful to close the detection loophole in quantum key distribution. Moreover, this protocol is based on the linear optics, which makes it feasible in current technology.
\end{abstract}
 \maketitle

\section{Introduction}
In the past few decades, a series of attractive achievements in the quantum information field have been made. Quantum computation and quantum communication will no longer be a dream \cite{book}. In  quantum communication, there are some important protocols have been proposed, such as quantum teleportation \cite{teleportation}, quantum key distribution \cite{key,key1}, quantum state sharing \cite{QSS}, quantum secure direct communication \cite{QSDC}, and so on \cite{densecoding,other1,other2,other3,other4}. Photon is the best candidate to carry and distribute quantum information, because it has fast transmission speed and is easy to control. Unfortunately, the unavoidable absorption and scattering in a transmission quantum channel places a serious limitation on the length of the communication distances \cite{loss}. The photon loss becomes one of the main obstacles in long-distance quantum communication. It not only decreases the efficiency of the communication, but also will make the communication insecure, because the detection loophole \cite{amplification1,amplification2,amplification3}.

During the past decades, people developed two powerful quantum technologies to resist the photon loss. The first quantum technology is the quantum repeaters \cite{loss, repeater,repeater2}. By dividing the whole channel into several segments,
they first generate the entanglement in each segment. Finally, by entanglement swapping, they can set up the entanglement in the whole distance. The second quantum technology  which will be detailed in this paper is the quantum state amplification. Though the quantum repeater can extend the entanglement between the adjacent segments, they still require to distribute entanglement in each short-distance segment. In this way, during the transmission, the photons will lose with some probability. Briefly speaking, the photon loss will cause the single photon $|1\rangle$ degrade to a mixed state as $\eta|1\rangle\langle1|+(1-\eta)|0\rangle\langle0|$, which means the photon may be completely lost in the probability of $1-\eta$.

 In 2009, Ralph and Lund first proposed the concept of the noiseless linear amplification (NLA) to distill the new mixed state with relatively high fidelity from the input mixed state with low fidelity \cite{NLA0}. Since then, various NLA protocols have been proposed \cite{NLA1,NLA2,NLAadd1,NLAadd2,NLA3,NLAadd3,NLA4,NLA5,NLA6,NLA7,NLA8,NLA9,NLA10}. Current NLA protocol can be divided into three groups. The first group focused on the single photon \cite{NLA1,NLA2,NLAadd1,NLAadd2}.
 The second group focused on the single-photon entanglement (SPE)  \cite{NLA3,NLAadd3,NLA4,NLA5,NLA6,NLA7}, for the SPE is the simplest entanglement form, but
it has important applications  in cryptography \cite{cryptography,cryptography1}, state engineering \cite{engineering}, tomography \cite{tomography,tomography1}, entanglement purification \cite{singlepurification1,singlepurification2}, entanglement concentration \cite{singleconcentration1,singleconcentration2,singleconcentration3,nonlocality,telescopes},
 The third group focused on the continuous variables systems \cite{NLA8,NLA9,NLA10}.
 For example,  in 2012, Osorio \emph{et al.} experimentally realized the heralded noiseless amplification for the single-photon state (SPS) with the help of the single-photon source and the linear optics \cite{NLA1}.  Kocsis \emph{et al.} demonstrated the heralded noiseless amplification of a photon polarization qubit \cite{NLA2}. Zhang \emph{et al.} also proposed an NLA protocol for protecting the SPE \cite{NLA5}.

So far, all the existing NLA protocols for the SPS and SPE can only be performed for one time. That is the fidelity of the initial state can be increased for one step. In the practical high noisy applications, the photon loss is usually high. In this way, after performing the amplification protocol, the quality of the entanglement may not reach the standard for secure and highly efficient long-distance quantum communication.  Because in order to close the detection loophole, the fidelity of the state is the higher the better. In this way, we should seek for the efficient approach to realize the amplification. In this paper, based on the linear optics, we propose an efficient  NLA protocol for protecting both the SPS and SPE, respectively.
In our protocol, the amplification is cascaded. That is the fidelity of the SPS and SPE can be increased step by step.

This paper is organized as follows: In Sec. II, we present the cascaded NLA protocol for SPS. In Sec. III, we present our cascaded amplification protocol for SPE. In Sec. IV, we present a discussion.  Finally, in Sec. V, we present a conclusion.

\section{Cascaded NLA protocol for SPS}

Before we start to describe our protocol, we first introduce the basic principle of the NLA protocol, based on the work of Gisin  \emph{et al.}\cite{NLA1}. It is composed of the variable fiber beam splitter (VBS) with the transmittance of $t$ and the 50:50 beam splitter (BS).  The schematic drawing is shown in Fig. 1. The  mixed state of the SPS can be described as
\begin{eqnarray}
\rho_{in}=\eta|1\rangle_{a1}\langle1|+(1-\eta)|0\rangle_{a1}\langle0|.\label{initial}
\end{eqnarray}

In Ref.\cite{NLA1}, an auxiliary single photon is required. The auxiliary photon passes through the VBS, which can generate a single-photon entangled state as
\begin{eqnarray}
|\varphi_{A}\rangle=\sqrt{t}|1\rangle_{b1}|0\rangle_{b2}+\sqrt{1-t}|0\rangle_{b1}|1\rangle_{b2}.
\end{eqnarray}
\begin{figure}[!h]
\begin{center}
\includegraphics[width=6cm,angle=0]{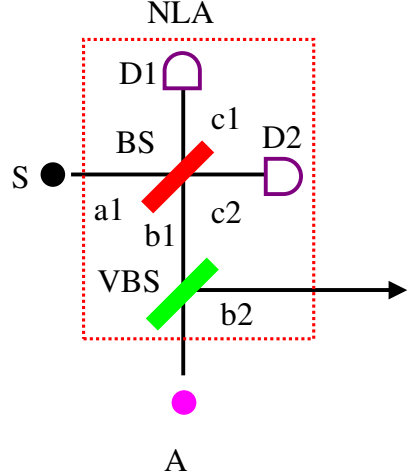}
\caption{The schematic drawing of the NLA protocol in Ref. \cite{NLA1} for SPS. S represents the single-photon source. VBS means the variable fiber beam splitter, and BS means the 50:50 beam splitter. By passing through the NLA unit, the output mixed state with higher fidelity can be distilled. }
\end{center}
\end{figure}
We make the photons in the a1 and b1 modes pass through the BS, which can make
\begin{eqnarray}
|1\rangle_{a1}=\frac{1}{\sqrt{2}}(|1\rangle_{c1}+|1\rangle_{c2}),\qquad
|1\rangle_{b1}=\frac{1}{\sqrt{2}}(|1\rangle_{c1}-|1\rangle_{c2}).
\end{eqnarray}
Then, the output photon in the c1 and c2 modes are detected by the single-photon detectors D1 and D2, respectively. We select the cases which make D1 or D2 detects exactly one photon. In this way, we can finally obtain a new mixed state as
\begin{eqnarray}
\rho_{out1}=\eta_{1}|1\rangle_{b2}\langle1|+(1-\eta_{1})|0\rangle_{b2}\langle0|,\label{new1}
\end{eqnarray}
where the coefficient $\eta_{1}$ is
\begin{eqnarray}
\eta_{1}=\frac{(1-t)\eta}{(1-t)\eta+(1-\eta)t}=\frac{(1-t)\eta}{\eta-2\eta t+t}.\label{f1}
\end{eqnarray}
The success probability of the protocol is
\begin{eqnarray}
P=(1-t)\eta+t(1-\eta)=t+\eta-2t\eta.
\end{eqnarray}

\begin{figure}[!h]
\begin{center}
\includegraphics[width=8cm,angle=0]{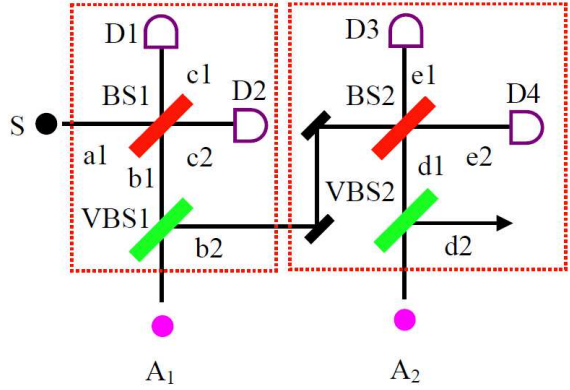}
\caption{The schematic drawing of the two-level NLA protocol for SPS. }
\end{center}
\end{figure}

We denote the amplification factor $g\equiv\frac{\eta_{1}}{\eta}$, so that we can obtain
\begin{eqnarray}
g=\frac{1-t}{\eta-2\eta t+t}.\label{g}
\end{eqnarray}
In order to realize the amplification, it is required that $g>1$ . It can be calculated that we can obtain $g>1$ only if $t<\frac{1}{2}$.

 In the section, we put forward an efficient cascaded amplification protocol for the SPS with the help of the NLA unit.
  We first describe the two-level cascaded NLA protocol, as described in Fig. 2. From Fig.2, we require two single photons $A_{1}$ and $A_{2}$ as auxiliary.
 After one of the single-photon detectors $D1$ or $D2$ registers one photon, the state in the spatial mode $b2$ must be an amplified mixed state of the form
 in Eq.(\ref{new1}). In this way, the new mixed state can be regarded as the initial state in the second amplification, with the help of another auxiliary single photon $A_{2}$. In the second round, we still choose the case that only one of the single-photon detectors $D3$ or $D4$ register the photon.
If the protocol is successful, we can obtain a new mixed state as
\begin{eqnarray}
\rho_{out2}=\eta_{2}|1\rangle_{d2}\langle1|+(1-\eta_{2})|0\rangle_{d2}\langle0|,\label{new2}
\end{eqnarray}
 with
 \begin{eqnarray}
 \eta_{2}&=&\frac{\eta_{1} (1-t)}{t+\eta_{1}-2\eta_{1} t}=\frac{\eta(1-t)^{2}}{t^{2}+\eta-2\eta t}.
 \end{eqnarray}

 It is straightforward to extend this two-level cascaded NLA protocol to the arbitrary level cascaded NLA protocol, as described in Fig. 3.
 In the protocol, we make the input photon pass through N NLA units, successively. The output photon state from the previous NLA unit enter the next NLA unit as the input one. The transmittance of the VBS in each NLA unit is required to meet $t<\frac{1}{2}$. With the help of N auxiliary photons, say A$_{1}$, A$_{2}$, $\cdots\cdots$ A$_{N}$,
 the initial SPS can be cascaded to amplify for N time.
\begin{figure}[!h]
\begin{center}
\includegraphics[width=8cm,angle=0]{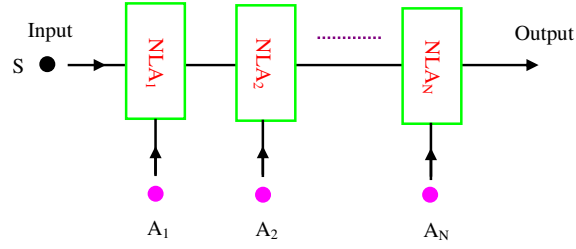}
\caption{The schematic drawing of our cascaded amplification protocol for SPS. We make the input photon state pass through N NLA units, successively. With the help of N auxiliary single photons, we can realize the cascade amplification for the SPS. }
\end{center}
\end{figure}
Based on the description in Sec. II, after N NLA unit, we can obtain the output new mixed photon state  as
\begin{eqnarray}
\rho_{out_{N}}=\eta_{N}|1\rangle\langle1|+(1-\eta_{N})|0\rangle\langle0|.\label{newn}
\end{eqnarray}
The $\eta_{N}$ is the fidelity of the output mixed state after NLA$_{N}$. We can calculate the fidelity of the mixed state after the photon pass through each NLA unit as
\begin{eqnarray}
\eta_{1}&=&\frac{\eta (1-t)}{t+\eta-2\eta t},\nonumber\\
\eta_{2}&=&\frac{\eta_{1} (1-t)}{t+\eta_{1}-2\eta_{1} t},\nonumber\\
&\cdots\cdots& \nonumber\\
\eta_{N}&=&\frac{\eta_{N-1} (1-t)}{t+\eta_{N-1}-2\eta_{N-1} t}.\label{fidelity1}
\end{eqnarray}

After NLA$_{N}$, we define the total amplification factor G as
\begin{eqnarray}
G\equiv\frac{\eta_{N}}{\eta}=\frac{\eta_{N}}{\eta_{N-1}}\frac{\eta_{N-1}}{\eta_{N-2}}\cdots\frac{\eta_{2}}{\eta_{1}}\frac{\eta_{1}}{\eta}=g_{1}g_{2}\cdots g_{N}.\label{factor}
\end{eqnarray}
Under the case that the transmittance of each VBS meet $t<\frac{1}{2}$, we can ensure arbitrary $g_{n}>1$. Therefore, increasing the number of NLA unit ($N$) can effectively increase the fidelity of the final output SPS. Especially, if $N\rightarrow\infty$, we can make $\eta_{N}\rightarrow1$.

Meanwhile, the success probability to distill the new mixed state after each NLA can be written as
\begin{eqnarray}
P_{1}&=&t+\eta-2t\eta,\nonumber\\
P_{2}&=&P_{1}[t+\eta_{1}-2\eta_{1}t],\nonumber\\
&\cdots\cdots&\nonumber\\
P_{N}&=&P_{N-1}[t+\eta_{N-1}-2\eta_{N-1}t].\label{P1}
\end{eqnarray}

Obviously, as arbitrary $P_{n}<1$, increasing the number of the NLA unit will reduce the success probability. Therefore, it is a trade-off between the fidelity and the success probability. In order to obtain the SPS with high fidelity, we need to consume large number of the input single photons.

\section{Cascade amplification for the SPE}
 In 2012, the setup of NLA described in Ref. \cite{NLA1} was developed to amplify SPE \cite{NLA2}, which is shown in Fig. 2. Briefly speaking, a single-photon source S emits single-photon entangled state, which is in the spatial mode $a1$ and $b1$. The form of SPE can be written as
\begin{eqnarray}
|\phi\rangle_{AB}=\frac{1}{\sqrt{2}}(|10\rangle_{AB}+|01\rangle_{AB}).
\end{eqnarray}
Due to the photon loss, the SPE degrades to a mixed state as
\begin{eqnarray}
\rho_{in}=\eta'|\phi\rangle_{AB}\langle\phi|+(1-\eta')|vac\rangle\langle vac|.\label{entangle}
\end{eqnarray}

\begin{figure}[!h]
\begin{center}
\includegraphics[width=10cm,angle=0]{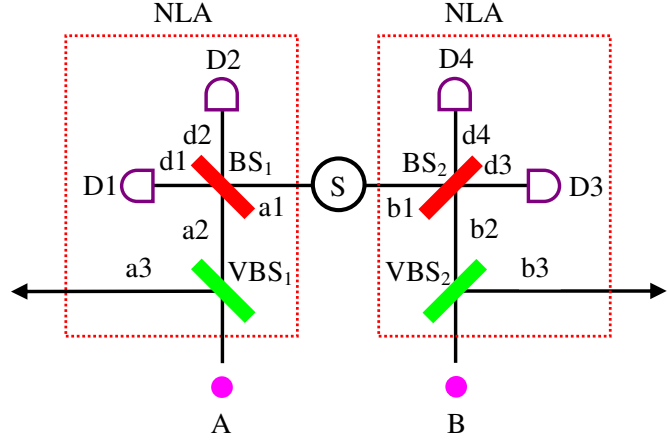}
\caption{The schematic drawing of the NLA protocol in Ref. \cite{NLA5} for the SPE. In the protocol, both the two parties makes the photons in their hands pass through a NLA unit, simultaneously. With the help of two auxiliary single photons, we can distill the new mixed state with higher fidelity.}
\end{center}
\end{figure}

For realizing the amplification, two auxiliary photons are required and both the two parties need to run the same operation as described above, simultaneously. After the amplification, we can obtain the new mixed state as
\begin{eqnarray}
\rho_{out2}=\eta_{1}'|\phi\rangle_{AB}\langle\phi|+(1-\eta_{1}')|vac\rangle\langle vac|,
\end{eqnarray}
with the success probability of
\begin{eqnarray}
P'=\eta' t(1-t)+(1-\eta')t^{2}=(1-2\eta')t^{2}+\eta' t.
\end{eqnarray}
The fidelity of the new mixed state is
\begin{eqnarray}
\eta'_{1}=\frac{\eta' (1-t)}{\eta' (1-t)+(1-\eta')t}=\frac{\eta' (1-t)}{\eta'-2\eta't+t}.\label{f2}
\end{eqnarray}
Interestingly, it can be found that the form of the fidelity in Eq. (\ref{f2}) is the same as that in Eq. (\ref{f1}). Therefore, we can also obtain that when the transmittance of each VBS meets $t<\frac{1}{2}$, the amplification factor $g>1$.

\begin{figure}[!h]
\begin{center}
\includegraphics[width=12cm,angle=0]{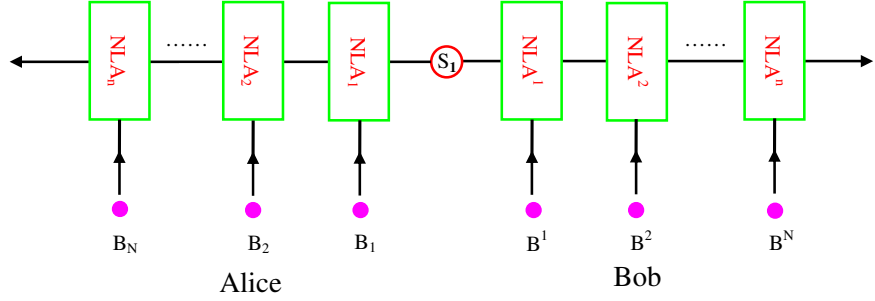}
\caption{The schematic drawing of our cascaded amplification protocol for SPE. Each of the two parties makes the photon in his/her hand pass through N NLA units, successively. With the help of 2N auxiliary single photons, we can realize the cascade amplification for the single-photon entangled state. }
\end{center}
\end{figure}
In the section, the NLA unit will be used to realize the cascade amplification for the SPE state. As shown in Fig. 5, due to the environmental noise, Alice and Bob share a mixed state as Eq. (\ref{entangle}). For realizing the amplification, each of them needs to prepare N NLA units, say NLA$_{1}$, NLA$_{2}$,$\cdots\cdots$ NLA$_{N}$, and NLA$^{1}$, NLA$^{2}$, $\cdots\cdots$ NLA$^{N}$, respectively. Each of Alice and Bob makes the single photon in his/her hand pass through the N units, simultaneously. Similarly, during the amplification process, each of them needs to introduce N auxiliary photons, say B$_{1}$, B$_{2}$, $\cdots\cdots$ B$_{N}$, and B$^{1}$, B$^{2}$, $\cdots\cdots$ B$^{N}$. After the N NLA units, we can obtain the final output quantum state as
\begin{eqnarray}
\rho'_{out}=\eta_{N}'|\phi\rangle_{AB}\langle\phi\rangle_{AB}|+(1-\eta_{N}')|vac\rangle\langle vac|,
\end{eqnarray}
where $\eta_{N}'$ is the fidelity of the mixed state when both the input photons from Alice and Bob pass through N NLA units.

We can also calculate the fidelity and success  possibility of the cascaded amplification protocol for the SPE state. The fidelity of the mixed state when both the two parties make the input photon pass through arbitrary N NLA units can be written as
\begin{eqnarray}
\eta_{1}'&=&\frac{\eta' (1-t)}{\eta' (1-t)+(1-\eta')t},\nonumber\\
\eta_{2}'&=&\frac{\eta_{1}' (1-t)}{\eta_{1}' (1-t)+(1-\eta_{1}')t},\nonumber\\
&\cdots\cdots& \nonumber\\
\eta_{N}'&=&\frac{\eta_{N-1}' (1-t)}{\eta_{N-1}' (1-t)+(1-\eta_{N-1}')t},\label{fidelity2}
\end{eqnarray}
where the subscript "1", "2", $\cdots\cdots$ "N" mean the number of the NLA units adopted by each of the two parties. Similarly, it can be found Eq. (\ref{fidelity2}) has the same form as Eq. (\ref{fidelity1}).

Therefore, when both the two parties make their photon pass through N NLA units, the total amplification factor $G$ is the same as Eq. (\ref{factor}).
In this way, under the case that $t<\frac{1}{2}$, we can effectively increase the amplification factor by increasing the number of the NLA units.

Certainly, we can also calculate the success probability of the cascaded amplification protocol as
\begin{eqnarray}
P_{1}'&=&(1-2\eta')t^{2}+\eta' t,\nonumber\\
P_{2}'&=&P_{1}'[(1-2\eta_{1}')t^{2}+\eta_{1}' t],\nonumber\\
&\cdots\cdots&\nonumber\\
P_{N}'&=&P_{N-1}'[(1-2\eta_{N-1}')t^{2}+\eta_{N-1}' t],\label{P2}
\end{eqnarray}
where the subscript "1", "2", $\cdots$ "N" are the number of the NLA units used in each of the two parties. Similarly, increasing the fidelity will also sacrifice the success probability. For obtaining the SPE with high fidelity, we still need to consume large amount of initial input state.

\section{Discussion}
So far, based on the work of Refs.\cite{NLA1,NLA5}, we have fully described our cascaded NLA protocol for both SPS and SPE.
The NLA unit which is composed of the VBS and BS, is the key element of the two protocols. In our protocol,
 we make the target photon pass through N NLA units, successively. Under the case that the transmittance of the VBS ($t$) meets $t<\frac{1}{2}$,
 when the target photon pass through a NLA unit, we can realize an amplification with the help of an auxiliary photon.
 In this way, by making the photon pass through N NLA units successively, we can finally realize N cascaded amplification.\\

\begin{figure}[!h]
\begin{center}
\includegraphics[width=15cm,angle=0]{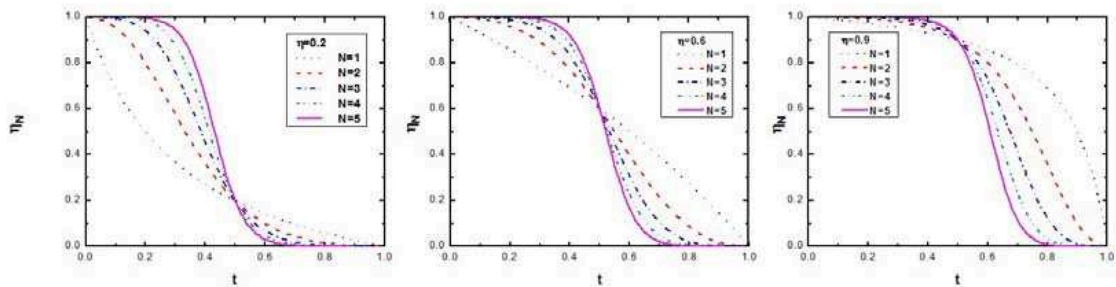}
\caption{The fidelity ($\eta_{N}$) of the distilled new mixed state as a function of the transmittance ($t$) of the VBS, when N (2N) NLA units were used to cascade amplify the SPS (SPE). For comparison,  we make N=1, 2, 3, 4 and 5, respectively. Meanwhile, we suppose that both the two protocols are operated under low initial fidelity ($\eta=0.2$), medium initial fidelity ($\eta=0.6$), and high initial fidelity ($\eta=0.9$). }
\end{center}
\end{figure}
According to Eq. (\ref{fidelity1}) and Eq. (\ref{fidelity2}), the fidelity ($\eta_{N}$) of the output mixed states largely depends on the initial fidelity ($\eta$) of the input state, the transmittance ($t$) of the VBS, and the number ($N$) of the NLA unit. Fig. 6 shows $\eta_{N}$ as a function of $t$. For comparison, we suppose the protocols are operated under low initial fidelity ($\eta=0.2$), medium initial fidelity ($\eta=0.6$) and high initial fidelity ($\eta=0.9$), respectively, and make N=1, 2, 3, 4, and 5. The fidelity curve of the protocols in Ref. \cite{NLA1,NLA2} is the same as that for $N=1$. It can be found that $\eta_{N}$ reduces with the growth of $t$. The five curves in each figure interact at the point of $t=0.5$. Under the case that $t=0.5$, $\eta_{N}=\eta$, that is, the output mixed state is the same as the input state. Actually, when $t=0.5$, the VBS become the BS, the whole amplification process is converted to the standard  teleportation process.
On the other hand, increasing the number of the NLA unit can effectively increase the fidelity of the output states, especially under low initial fidelity condition. In practical applications, the photon loss is usually high. Fig. 7 shows the the fidelity $\eta_{N}$ altered with $N$ under practical high photon loss condition ($\eta=0.2$). It can be found that when $t=0.2$, $\eta_{1}$ is only 0.5,
while $\eta_{5}$ can reach 0.996. Therefore, by selecting the suitable VBS and the suitable value of N, we can make $\eta_{N}\rightarrow1$.
In this way, our protocols may provide an effectively way to close the detection loophole in QKD \cite{NLA9}. Certainly, we should point out that though the $\eta_{N}$ increases with $N$, we cannot reach $\eta_{N}=1$. The limitation of $\lim_{N\rightarrow\infty}\eta_{N}=1$. That is the $\eta_{N}=1$ is a fixed point for the iterative equations in both Eq. (\ref{fidelity1}) and Eq. (\ref{fidelity2}).
On the other hand, under high photon loss condition ($\eta=0.2$), the protocols in Refs. \cite{NLA1,NLA2} can obtain relatively high fidelity only under the extreme condition that $t\rightarrow0$. For example, when $N=1$, $\eta_{1}$ is 0.69 under $t=0.1$, and $\eta_{1}$ can reach 0.96 under $t=0.01$. In current experimental conditions, the VBS with $t\rightarrow 0$ is unavailable. However, with the growth of $N$, the requirement for $t$ is largely reduced. For example, when $N=5$, $\eta_{5}$ can reach 0.996 under $t=0.2$. Therefore, it is much easier for us to obtain high fidelity under practical experimental conditions.

\begin{figure}[!h]
\begin{center}
\includegraphics[width=6cm,angle=0]{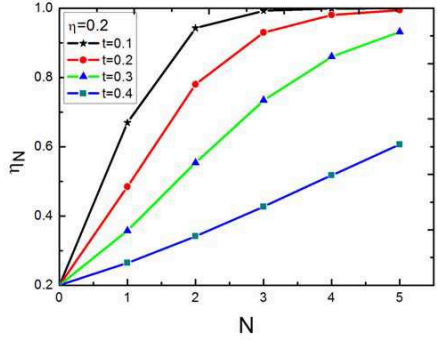}
\caption{ The fidelity $\eta_{N}$ altered with the iteration number $N$. Here we let $N=1, 2, 3, 4, 5$ and the initial fidelity $\eta=0.2$.}
\end{center}
\end{figure}
Finally, we will discuss the experimental realization of our protocol. The VBS and BS are the key elements. Ref. \cite{NLA1} reported the experimental results about the NLA with the help of the VBS. The protocol can increase the probability $\eta_{t}$ of the single photon $|1\rangle$ from a mixed state $\eta_{t}|1\rangle\langle1|+(1-\eta_{t})|0\rangle\langle0|$ by adjusting the splitting ratio of VBS from
 50:50 to 90:10. Based on the experiments, our requirement for $t<1/2$ can be easily realized under current technology. In our protocol, the processing of the photons passing through the BS is essentially the Hong-Ou-Mandel (HOM) interference, so that the two photons should be indistinguishable in every degree of freedom. Ref. \cite{NLA1} also measured the HOM interference on each BS. Their experimental results for each BS are 93.4 $\pm $5.9\% and 92.1 $\pm$ 5.7\%, respectively. In Fig. 8, we design  the possible experimental realization of the two-level amplification for the SPS. Based on Ref. \cite{Pan,Pan1,shengpra,shengpra2}, with the help of spontaneous parametric down-conversion (SPDC) source,  we make a pump pulse of
ultraviolet light pass through a beta barium borate (BBO) crystal and produce correlated pairs of photons into the
modes a$_{1}$ and b$_{1}$. Then it is reflected on the mirror and traverses the crystal for a second time, and produces correlated pairs of photons into
the modes a$_{2}$ and b$_{2}$. The Hamiltonian can be approximately described as
 \begin{eqnarray}
H_{PDC}&=&|0\rangle+p(|a^{\dag}_{1}b^{\dag}_{1}\rangle+|a^{\dag}_{2}b^{\dag}_{2}\rangle)|0\rangle+p^{2}(|a^{\dag}_{1}b^{\dag}_{1}\rangle
\otimes|a^{\dag}_{2}b^{\dag}_{2}\rangle\nonumber\\
&+&|a^{\dag}_{1}b^{\dag}_{1}\rangle\otimes|a^{\dag}_{1}b^{\dag}_{1}\rangle+|a^{\dag}_{2}b^{\dag}_{2}\rangle\otimes|a^{\dag}_{2}b^{\dag}_{2}\rangle)|0\rangle+o(p).
\end{eqnarray}

\begin{figure}[!h]
\begin{center}
\includegraphics[width=10cm,angle=0]{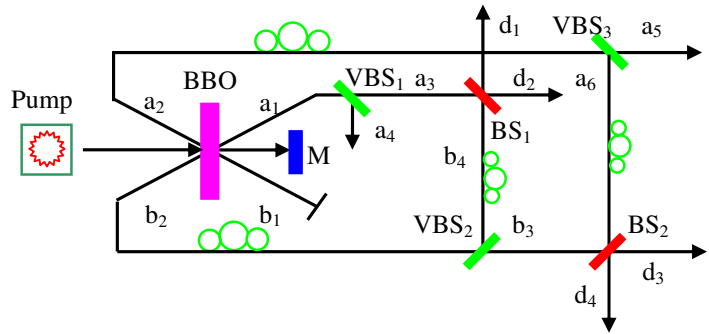}
\caption{The possible experimental realization of the two-level amplification for the SPS. We adopt the SPDC source to produce four photons each in the a$_{1}$, b$_{1}$, a$_{2}$, and b$_{2}$ mode. The photon in the a$_{1}$ mode is the target photon, which pass through the VBS$_{1}$ to generate the input mixed state. The photons in a$_{2}$ and b$_{2}$ modes are used as the auxiliary photons. }
\end{center}
\end{figure}
In the experiment, we select the item $(|a^{\dag}_{1}b^{\dag}_{1}\rangle
\otimes|a^{\dag}_{2}b^{\dag}_{2}\rangle)|0\rangle$ with the help of the coincidence measurement, which generates four photons in the modes a$_{1}$, b$_{1}$, c$_{1}$, and d$_{1}$, simultaneously. Under this case, we make the photon in a$_{1}$ mode pass through VBS$_{1}$ with the transmittance of t$_{1}$ to generate a mixed state in the a$_{3}$. If the SPS is reflected to the a$_{4}$ mode, it means that the photon is lost. The photon in the mode b$_{1}$ can be used to judge the single-photon in  a$_{1}$ mode. By changing  t$_{1}$, we can obtain the different mixed state.
The photon in both a$_{2}$ and b$_{2}$ mode are used as the auxiliary photons. After making it pass through VBS$_{2}$ with the transmittance $t_{2}<\frac{1}{2}$, we make the photons in the a$_{3}$ and b$_{4}$ modes pass through the BS$_{1}$ and detect the photon in the d$_{1}$ and d$_{2}$ modes. In this way, the first level of amplification  can be realized. Next, the photon in the a$_{2}$ mode is used as the second auxiliary photon. Similarly, we make it pass through VBS$_{3}$ with the transmittance $t_{2}<\frac{1}{2}$, and then make the photons in the a$_{6}$ and b$_{3}$ modes pass through the BS$_{2}$. We can finally realize the second amplification. In current technology, with the help of cascaded SPDC sources, the generation of eight-photon  entanglement has been realized \cite{eight1,eight2}. Such  cascaded SPDC sources can be used to implement the experiment for multi-level cascaded NLA protocol.

\section{conclusion}
In conclusion, we put forward an efficient cascaded amplification NLA protocol for both the SPS and SPE, respectively. In our protocol, we make target photon pass through several NLA units, successively. With the help of some auxiliary single photons, we can realize the cascaded amplification for both the SPS and SPE. This protocol is based on the linear optics, which is extremely suitable in current technology. In the discussion, we also design a possible realization with current SPDC source. The most advantage of this protocol is that the fidelity can be iterated to obtain a higher value. It provides us that this protocol is extremely useful in a large noisy channel, which may be used to close the detection loophole in current long-distance quantum communication.

\section{Acknowledgements}
This work is supported by the National Natural Science Foundation of
China under Grant Nos. 11474168 and 61401222, the Qing Lan Project in Jiangsu Province,  and the Priority Academic Program Development of Jiangsu Higher Education Institutions.

\end{document}